\documentstyle[preprint,aps,prl]{revtex}

\author{Sergey V. Kuplevakhsky\\
Department of Physics, Kharkov State University,\\
310077 Kharkov, Ukraine}
\title{Exact solution of the Lawrence-Doniach model in parallel magnetic
fields }

\begin{document}

\maketitle
\begin{abstract}
For the first time, we obtain the complete and exact analytical solution of
the Lawrence-Doniach model for layered superconductors in external parallel
magnetic fields. By solving a nontrivial mathematical problem of exact
minimization of the free-energy functional, we derive a closed,
self-consistent system of mean-field equations involving only two variables.
Exact solutions to these equations prove simultaneous penetration of
Josephson vortices into all the barriers, yield a completely new expression
for the lower critical field, refute the concept of a triangular Josephson
vortex lattice and clarify the physics of Fraunhofer oscillations of the
total critical Josephson current.

\bigskip\

\end{abstract}

In this paper, we obtain for the first time the complete and exact
analytical solution of the popular phenomenological Lawrence-Doniach\cite{LD}
(LD) model for layered superconductors in external parallel magnetic fields.

At present, there is a universal belief that the LD model can adequately
describe high-$T_c$ superconductors exhibiting the intrinsic Josephson
effect.\cite{KSKM92} Surprisingly, despite a large number of publications on
the LD model over the recent years, it has not been realized yet that in the
presence of a parallel magnetic field the LD free-energy functional provides
a rare example of exactly solvable models in theoretical physics. (As an
exception, there is a particular exact solution of Theodorakis,\cite{Th90}
valid in a restricted field range.) Moreover, even a closed, self-consistent
system of mean-field equations for the LD functional has not been obtained
up until now, which in some cases leads to spurious results. Thus,
calculations of the lower critical field\cite{B73,CCH91,K93} are based on an
arbitrary assumption of single Josephson vortex penetration and a continuum
approximation, incompatible with the discrete nature of the LD model. (This
treatment was criticized in Ref. 3.) The claim\cite{BCG92} that the
Fraunhofer pattern of the critical Josephson current occurs in the absence
of Josephson vortices is at odds with the well-known situation\cite{T96} in
a single junction. Furthermore, a hypothesis\cite{BC91} of a triangular
lattice of Josephson vortices stands in direct contradiction to the exact
solution of Ref. 3. Here we show that the origin of these inconsistencies is
in an incorrect mathematical approach to the minimization of the LD
functional, neglecting principal aspects of any gauge theory.\cite{SF88}
Based on exact variational methods, we derive a remarkably simple, closed,
self-consistent set of mean-field equations involving only two variables. As
these equations turn out to be a special limiting case of a recently
developed microscopic theory,\cite{K98} we concentrate here mostly on the
problem of exact minimization of the LD functional and provide a brief
summary of the main new physical results at the end of the paper.

We begin by reminding basic features of the LD model.\cite{LD,KLB75} In this
model, the temperature $T$ is assumed to be close to the ''intrinsic''
critical temperature $T_{c0}$ of individual layers:
\begin{equation}
\label{1}\tau \equiv \frac{T_{c0}-T}{T_{c0}}\ll 1.
\end{equation}
The superconducting (S) layers are assumed to have negligible thickness
compared to the ''intrinsic'' coherence length $\zeta (T)\propto \tau
^{-1/2} $, the penetration depth $\lambda (T)\propto \tau ^{-1/2}$, and the
layering period $p$. Taking the layering axis to be $x$, choosing the
direction of the external magnetic field ${\bf H}$ to be $z$ [${\bf H}%
=(0,0,H)$] and setting $\hbar =c=1$, we can write the LD free-energy
functional as

$$
\Omega _{LD}\left[ f_n,\phi _n,\frac{d\phi _n}{dy},A_x,A_y;H\right] =\frac{%
pH_c^2(T)}{4\pi }W_z{\int\limits_{L_{y1}}^{L_{y2}}}dy
{\sum _{n=-\infty }^{+\infty }}\left[ -f_n^2(y)+\frac
12f_n^4(y)\right.
$$

$$
+\zeta ^2(T)\left[ \frac{df_n(y)}{dy}\right] ^2+\zeta ^2(T)\left[ \frac{%
d\phi _n(y)}{dy}-2eA_y(np,y)\right] ^2f_n^2(y)
$$

$$
\left. +\frac{r(T)}2\left[ f_{n-1}^2(y)+f_n^2(y)-2f_n(y)f_{n-1}(y)\cos \Phi
_{n,n-1}(y)\right] \right]
$$

\begin{equation}
\label{2}\left. +\frac{4e^2\zeta ^2(T)\lambda ^2(T)}p
{\int\limits_{(n-1)p}^{np}}dx\left[ \frac{\partial A_y(x,y)}{\partial x}
-\frac{\partial A_x(x,y)}{\partial y}-H\right] ^2\right] ,
\end{equation}

$$
\Phi _{n,n-1}(y)=\phi _n(y)-\phi _{n-1}(y)-2e
{\int\limits_{(n-1)p}^{np}}dxA_x(x,y),
$$
Here ${\bf A}=(A_x,A_y,0)$ is the vector potential, continuous at the
S-layers: ${\bf A(}np-0,y)={\bf A(}np+0,y)={\bf A(}np,y)$; $W_z$ is the
length of the system in the $z$ direction; $f_n(y)$ [$0\leq f_n(y)\leq 1$]
and $\phi _n(y)$ are, respectively, the reduced modulus and the phase of the
pair potential $\Delta _n(y)$ in the $n$th superconducting layer:
\begin{equation}
\label{3}\Delta _n(y)=\Delta (T)f_n(y)\exp \phi _n(y),
\end{equation}
with $\Delta (T)$ being the ''intrinsic'' gap [$\Delta (T)\propto \tau
^{1/2} $]; $H_c(T)$ is the thermodynamic critical field; $r(T)=2\alpha
_{ph}\tau ^{-1}$ is a dimensionless phenomenological parameter of the
Josephson interlayer coupling ($0<\alpha _{ph}\ll 1$). The local magnetic
field ${\bf h}=(0,0,h)$ obeys the relation
\begin{equation}
\label{4}h(x,y)=\frac{\partial A_y(x,y)}{\partial x}-\frac{\partial A_x(x,y)
}{\partial y}.
\end{equation}

Our task now is to establish a closed, complete, self-consistent system of
mean-field equations of the theory, which is mathematically equivalent to
the minimization of (\ref{2}) with respect to $f_n$, $\phi _n$, and ${\bf A}$%
. First, we want to point out a common mistake\cite{BLK92} in the approach
to this problem: It has not been realized in the literature that variations
with respect to $\phi _n$ and ${\bf A}$ are not independent and do not yield
a complete set of equations. Indeed, as the functional (\ref{2}) is
invariant under the gauge transformations
\begin{equation}
\label{5}\phi _n(y)\rightarrow \phi _n(y)+2e\lambda (np,y),\text{ }%
A_i(x,y)\rightarrow A_i(x,y)+\partial _i\lambda (x,y),\text{ }i=x,y,
\end{equation}
where $\lambda (x,y)$ is an arbitrary smooth function of $x$, $y$ in the
whole region $\left( -\infty <x<+\infty \right) \times \left(
L_{y1}<y<L_{y2}\right) $, variational derivatives with respect to $\phi _n$,
and $A_{x\text{, }}A_y$ are related by the fundamental identities
\begin{equation}
\label{6}2e\frac{\delta \Omega _{LD}}{\delta \phi _n(y)}\equiv \frac
\partial {\partial y}\frac{\delta \Omega _{LD}}{\delta A_y(np,y)}+\frac{%
\delta \Omega _{LD}}{\delta A_x(np+0,y)}-\frac{\delta \Omega _{LD}}{\delta
A_x(np-0,y)}.
\end{equation}
Being a consequence of Noether's second theorem, such identities are typical
of any gauge theory.\cite{SF88} They imply that the number of independent
Euler-Lagrange equations is less than the number of variables, and
complementary relations should be imposed to eliminate irrelevant degrees of
freedom and close the system mathematically. Whereas in bulk superconductors
and single junctions the elimination of unphysical degrees of freedom is
accomplished by fixing the gauge, in periodic weakly coupled structures this
problem has additional implications.\cite{K98} Namely, in the presence of
the Josephson interlayer coupling the quantities $\Phi _{n,n-1}$ are not
independent but subject to a set of constraint relations. Unfortunately,
this fundamental feature was not noticed in any previous publications on the
LD model.

Varying with respect to $A_x$, $A_y$ in the regions $(n-1)p<x<np$ under the
assumption $\delta A_x(x,L_{y1})=\delta A_x(x,L_{y2})=0$ yields
\begin{equation}
\label{9}\frac{\partial h(x,y)}{\partial y}=4\pi j_{n,n-1}(y)\equiv 4\pi
j_0f_n(y)f_{n-1}(y)\sin \Phi _{n,n-1}(y),
\end{equation}

\begin{equation}
\label{10}\frac{\partial h(x,y)}{\partial x}=0,
\end{equation}
where $j_{n,n-1}(y)$ is the density of the Josephson current between the $%
(n-1)$th and the $n$th layers, $j_0=r(T)p/16\pi e\zeta ^2(T)\lambda ^2(T)$.
Minimization with respect to $A_y(np,y)$ leads to boundary conditions at the
S-layers:
\begin{equation}
\label{11}h(np-0,y)-h(np+0,y)=\frac{pf_n^2(y)}{2e\lambda ^2(T)}\left[ \frac{%
d\phi _n(y)}{dy}-2eA_y(np,y)\right] .
\end{equation}

Equations (\ref{9})-(\ref{11}) should be complemented by boundary conditions
at the outer interfaces $y=L_{y1,}L_{y2}$. As we do not consider here
externally applied currents in the $y$ direction, the first set of boundary
conditions follows from the requirement $\left[ j_{ny}\right]
_{y=L_{y1,}L_{y2}}=0$:
\begin{equation}
\label{12}\left[ \frac{\partial \phi _n(x,y)}{\partial y}-2eA_y(x,y)\right]
_{y=L_{y1},L_{y2}}=0.
\end{equation}
Applied to Eqs. (\ref{11}), these boundary conditions show that the local
magnetic field at the outer interfaces is independent of the coordinate $x$:
$h(x,L_{y1})=h(L_{y1})$, $h(x,L_{y2})=h(L_{y2})$. The boundary conditions
imposed on $h$ should be compatible with Ampere's law $h(L_{y2})-h(L_{y1})=4%
\pi I$ obtained by integration of Eqs. (\ref{9}) over $y$, where
\begin{equation}
\label{13}I\equiv {\int\limits_{L_{y1}}^{L_{y2}}}
dyj_{n+1,n}(y)={\int\limits_{L_{y1}}^{L_{y2}}}dyj_{n,n-1}(y)
\end{equation}
is the total current in the $x$ direction.

Differentiating (\ref{11}) with respect to $y$ and employing (\ref{9}), we
arrive at the current-continuity equations for the S-layers:
$$
\frac \partial {\partial y}\left[ f_n^2(y)\left[ \frac{d\phi _n(y)}{dy}%
-2eA_y(np,y)\right] \right]
$$
\begin{equation}
\label{15}=\frac{r(T)}{2\zeta ^2(T)}f_n(y)\left[ f_{n-1}(y)\sin \Phi
_{n,n-1}(y)-f_{n+1}(y)\sin \Phi _{n+1,n}(y)\right] .
\end{equation}
Adding Eqs. (\ref{15}), integrating and using boundary conditions (\ref{12}%
), we get the first integral
\begin{equation}
\label{16}{\sum _{n=-\infty }^{+\infty }}f_n^2(y)\left[
\frac{d\phi _n(y)}{dy}-2eA_y(np,y)\right] =0.
\end{equation}
This equation has mathematical form of a constraint relation and states that
the total current in the $y$ direction is equal to zero.

The Euler-Lagrange equations for $\phi _n$ do not yield anything new and
only reproduce Eqs. (\ref{15}), as expected by virtue of Noether's
identities (\ref{6}). To obtain complementary constraint relations, closing
the system of the Euler-Lagrange equations and minimizing the free energy,
we must modify the variational procedure.

Noting that the kinetic energy of the intralayer currents in (\ref{2}) can
be minimized independently of the Josephson term, we impose additional
constraints
\begin{equation}
\label{18}\left[ \frac{\partial \phi _n(y)}{\partial y}-2eA_y(np,y)\right]
=0,
\end{equation}
compatible with boundary conditions (\ref{12}) and constraint relation (\ref
{16}). The requirement of compatibility with the current-conservation law (%
\ref{15}) automatically yields another set of constraints
\begin{equation}
\label{19}f_{n-1}(y)\sin \Phi _{n,n-1}(y)=f_{n+1}(y)\sin \Phi _{n+1,n}(y).
\end{equation}
The physical meaning of Eqs. (\ref{18}) and (\ref{19}) that provide the
sought necessary conditions for the true minimum of the free-energy
functional (\ref{2}) is obvious. Constraints (\ref{18}) minimize the kinetic
energy of the intralayer currents (it proves to be identically equal to
zero) and assure the continuity of the local magnetic field at the S-layers:
$h(np+0,y)=h(np-0,y).$ [See Eq. (\ref{11})]. These constraints appear
already in the case of decoupled S-layers. On the other hand, constraints (%
\ref{19}) are uniquely imposed by the Josephson interlayer coupling. Their
function is to make the Josephson energy stationary with respect to
variations of $\phi _n$ and to assure the continuity of the Josephson
current at the S-layers.

As no other conditions are imposed on the variables, we can satisfy (\ref{19}%
) by choosing
\begin{equation}
\label{20}f_n(y)=f_{n-1}(y)=f(y),\text{ }\Phi _{n+1,n}(y)=\Phi
_{n,n-1}(y)=\Phi (y).
\end{equation}
The establishment of constraints (\ref{18})-(\ref{20}), minimizing the free
energy and closing the set of mean-field equations, is a key result of this
paper. For example, these constraints automatically rule out any possibility
of previously proposed\cite{B73} single Josephson vortex penetration and the
hypothesis\cite{BC91} of a triangular Josephson vortex lattice. It should be
noted, however, that both the exact solution of Theodorakis\cite{Th90} for
the dense vortex state and early calculations\cite{LD,KLB75,T96} of the
upper critical field are fully compatible with relations (\ref{18})-(\ref{20}%
).

The remaining unphysical degree of freedom, related to the gauge invariance,
is eliminated by fixing the gauge:
\begin{equation}
\label{21}A_x(x,y)=0,\text{ }A_y(x,y)\equiv A(x,y).
\end{equation}
[Note that $\partial $$A/\partial x$ and $\partial ^2$$A/\partial x\partial
y $ are continuous at the S-layers by virtue of (\ref{11}), (\ref{18}), and (%
\ref{9}), (\ref{19}).] The second set of relations (\ref{20}) now yields $%
\phi _n(y)=n\phi (y)+\eta (y)$, where $\phi (y)$ is the coherent phase
difference (the same at all the barriers), and $\eta (y)$ is an arbitrary
function of $y$ that can be set equal to zero without any loss of generality.

From (\ref{10}), using the continuity conditions for $A,$ $\partial
A/\partial x$ and constraints (\ref{18}), we obtain $A(x,y)=\frac 1{2ep}
\frac{d\phi (y)}{dy}x,$ while the functional (\ref{2}) becomes
$$
\Omega _{LD}\left[ f,\phi ;H\right] =\frac{H_c^2(T)}{4\pi }W_xW_z
{\int\limits_{L_{y1}}^{L_{y2}}}dy\left[ -f^2(y)+\frac 12f^4(y)+\zeta
^2(T)\left[ \frac{df(y)}{dy}\right] ^2\right.
$$
\begin{equation}
\label{30}\left. +r(T)\left[ 1-\cos \phi (y)\right] f^2(y)+4e^2\zeta
^2(T)\lambda ^2(T)\left[ \frac 1{2ep}\frac{d\phi (y)}{dy}-H\right] ^2\right]
,
\end{equation}
where $W_x=L_{x2}-L_{x1}$. Minimizing (\ref{30}) with respect to $f(y)$
[with arbitrary $\delta $$f(L_{y1})$, $\delta $$f(L_{y2})$] and $\phi (y)$
[with $\delta \phi (L_{y1})=\delta \phi (L_{y2})=0$], we arrive at the
desired closed, self-consistent set of mean-field equations
\begin{equation}
\label{22}\Delta _n(y)=\Delta f(y)\exp \left[ in\phi (y)\right] ,
\end{equation}
\begin{equation}
\label{23}f(y)+\zeta ^2(T)\frac{d^2f(y)}{dy^2}-f^3(y)-r(T)\left[ 1-\cos \phi
(y)\right] f(y)=0,
\end{equation}
\begin{equation}
\label{24}\frac{df}{dy}(L_{y1})=\frac{df}{dy}(L_{y2})=0,
\end{equation}
\begin{equation}
\label{25}\frac{d^2\phi (y)}{dy^2}=\frac{f^2(y)}{\lambda _J^2}\sin \phi (y),
\end{equation}
\begin{equation}
\label{26}\lambda _J=\left( 8\pi ej_0p\right) ^{-1/2},
\end{equation}
\begin{equation}
\label{27}h(y)=\frac 1{2ep}\frac{d\phi (y)}{dy},
\end{equation}
\begin{equation}
\label{28}j(y)\equiv j_{n,n-1}(y)\equiv j_0f^2(y)\sin \phi (y)=\frac 1{4\pi
} \frac{dh(y)}{dy}
\end{equation}
that should be complemented by appropriate boundary conditions on $h(y)$
(see above) with $I\equiv
{\int\limits_{L_{y1}}^{L_{y2}}}dyj(y)$
, where $j(y)$ is the density of the Josephson current.

Remarkably, the coherent phase difference $\phi $ (the same for all the
barriers) obeys only one nonlinear second-order differential equation (\ref
{25}) with only one length scale, the Josephson penetration depth $\lambda
_J $ [Eq. (\ref{26})], as in the case of the Ferrell-Prange equation for a
single junction.\cite{BP82} [Mathematically, equation (\ref{25}) is a
solvability condition for the Maxwell equations.] Due to the factor $f^2$,
equation (\ref{25}) is coupled to nonlinear second-order differential
equation (\ref{23}) describing the spatial dependence of the superconducting
order parameter $f$ (the same for all the S-layers). Equations (\ref{24})
constitute boundary conditions for (\ref{23}). The Maxwell equations (\ref
{27}), (\ref{28}), combined together, yield Eq. (\ref{25}), as they should
by virtue of self-consistency.

It is instructive to compare the above equations with those of previous
publications, based on an incomplete minimization procedure. Thus, for $\Phi
_{n+1,n}(y)$ one introduces\cite{BC91,BCG92,BLK92} an infinite
non-self-consistent set of the so-called ''difference-differential''
equations, containing two length scales. By virtue of the constraint
relations (\ref{20}), in the gauge (\ref{21}) this set reduces to only one
equation (\ref{25}) with $f_n(y)=1$, while the second length scale, $\sqrt{2}%
\zeta (T)/\sqrt{r(T)}$, related to unphysical degrees of freedom, disappears
from the theory.

On the other hand, equations (\ref{22})-(\ref{28}) are only a limiting case
of the true microscopic equations,\cite{K98} if one identifies $r(T)$ with
the microscopic parameter $\alpha \zeta ^2(T)/a\xi _0$ and sets $a/p=0$,
where $\xi _0$ is the BCS coherence length, $a$ is the S-layer thickness [$%
\xi _0\ll a\ll \zeta (T),\lambda (T)$], and $\alpha =\frac{3\pi ^2}{7\zeta
(3)}{\int\limits_0^1}dttD(t)\ll 1$ [$D(t)$ is the
tunneling probability of the barrier between two successive S-layers].

Equations (\ref{23})-(\ref{28}) admit exact analytical solutions for all
physical situations of interest. Aside from the region near the second-order
phase transition to the normal state (because of the unphysical assumption
of negligible S-layer thickness, the LD model does not adequately describe
this regime\cite{T96}), these solutions stand in a one-to-one correspondence
with those of the microscopic theory.\cite{K98} For this reason, we only
briefly summarize the main physical results here, accentuating differences
between the exact solutions and previous non-self-consistent calculations.

The local magnetic field is independent of the coordinate in the layering
direction. [See Eq. (\ref{27}).] The Meissner phase in semi-infinite (along
the layers) samples persists up to the superheating field $H_s=(ep\lambda
_J)^{-1}.$ Contrary to previous suggestions,\cite{B73} Josephson vortices
penetrate all the barriers simultaneously and coherently, forming peculiar
structures that we term\cite{K98} ''vortex planes''. The existence of a
single vortex plane in an infinite (along the layers) sample becomes
energetically favorable at the lower critical field $H_{c1\infty }=2(\pi
ep\lambda _J)^{-1}$. (Previous calculations\cite{B73,CCH91,K93} of $%
H_{c1\infty }$, based on an invalid assumption of single Josephson vortex
penetration and an anisotropic continuum approximation, are incorrect.) In
the fields $H_{c1\infty }\ll H\ll [ep\zeta (T)]^{-1},$ with $r(T)\ll 1,$
equations (\ref{23})-(\ref{28}) reproduce the vortex-state solution of
Theodorakis.\cite{Th90} (The triangular vortex lattice\cite{BC91,I95} is not
allowed by the exact equations.) The magnetization in the vortex state
exhibits distinctive oscillatory behavior and jumps as a result of
vortex-plane penetration. For a certain field range, our calculations yield
a small paramagnetic effect. In contrast to previous assertions,\cite{BCG92}
the Fraunhofer pattern for the total critical Josephson current in layered
superconductors with $W\ll \lambda _J$ ($W=L_{y2}-L_{y1}$) occurs due to
successive penetration of the vortex planes and their pinning by the edges
of the sample. The first zero of the Fraunhofer pattern corresponds to the
lower critical field $H_{c1W}=\pi /epW$ of a finite sample. Finally, for the
upper critical field $H_{c2\infty }$ in an infinite layered superconductor,
equations (\ref{23})-(\ref{28}) yield the well-known results,\cite
{LD,KLB75,T96} as expected. We conclude by observing that the established
relation to the microscopic theory\cite{K98} casts light on the exact domain
of validity of the LD model.

\end{document}